%% file: ms.tex
\documentclass[12pt,preprint]{aastex}

\shorttitle{X-ray spectroscopy of BD+30$^{\circ}$3639}
\shortauthors{YU et al.}

\begin{document}


\title{The X-Ray Spectrum of a Planetary Nebula at High Resolution:
Chandra Gratings Spectroscopy of  BD+30$^{\circ}$3639}


\author{Young Sam Yu\altaffilmark{1}, Raanan Nordon\altaffilmark{2}, Joel H. Kastner\altaffilmark{1,3}, John Houck\altaffilmark{4}, Ehud Behar\altaffilmark{2,5}, Noam Soker\altaffilmark{2}}

\affil{1. Center for Imaging Science, Rochester Institute of Technology, Rochester, NY 14623-5604}
\affil{2. Department of Physics, Technion-Israel Institute of Technology, Haifa 32000, Israel}
\affil{3. Laboratoire d'Astrophysique de Grenoble, Universit\'e Joseph Fourier --- CNRS, BP 53, 38041 Grenoble Cedex, France}
\affil{4. Kavli Institute, Massachusetts Institute of Technology, Cambridge, MA 02139}
\affil{5. Senior NPP Fellow, Code 662, NASA/Goddard Space Flight Center, Greenbelt, MD 20771}

\begin{abstract}
  We present the results of the first X-ray gratings spectroscopy
  observations of a planetary nebula (PN), the X-ray-bright, young
  BD+30$^{\circ}$3639. We observed BD+30$^{\circ}$3639 for a total of
  300 ks with the Chandra X-ray Observatory's Low Energy Transmission
  Gratings in combination with its Advanced CCD Imaging Spectrometer
  (LETG/ACIS-S). The LETG/ACIS-S spectrum of BD+30$^{\circ}$3639 is
  dominated by H-like resonance lines of O {\sc viii} and C {\sc vi}
  and the He-like triplet line complexes of Ne {\sc ix} and O {\sc
    vii}. Other H-like resonance lines, such as N {\sc vii}, as well
  as lines of highly ionized Fe, are weak or absent. Continuum
  emission is evident over the range 6--18 \AA.  Spectral modeling
  indicates the presence of a range of plasma
  temperatures from T$_{x}$ ${\sim}$ 1.7 ${\times}$ 10$^{6}$ K to 2.9
  ${\times}$ 10$^{6}$ K and an intervening absorbing column
  $N_H\sim2.4\times10^{21}$ cm$^{-2}$. The same modeling conclusively
  demonstrates that C and Ne are highly enhanced, with abundance
  ratios of C/O ${\sim}$15--45 and Ne/O ${\sim}$3.3--5.0 (90\%
  confidence ranges, relative to the solar ratios), while N and Fe are
  depleted, N/O ${\sim}$0.0--1.0 and Fe/O ${\sim}$0.1--0.4. The
  intrinsic luminosity of the X-ray source determined from the
  modeling and the measured flux ($F_X = 4.1\times10^{-13}$ ergs
  cm$^{-2}$ s$^{-1}$) is $L_X\sim8.6\times10^{32}$ erg s$^{-1}$
  (assuming $D = 1.2$ kpc).

  These gratings spectroscopy results are generally consistent with
  earlier results obtained from X-ray CCD imaging spectroscopy of
  BD+30$^{\circ}$3639, but are far more precise. Hence the
  Chandra/LETGS results for BD+30$^{\circ}$3639 place severe new
  constraints on models of PN wind-wind interactions in which
    X-ray emitting gas within PNs is generated via shocks and the
    plasma temperature is moderated by effects such as heat conduction
    or rapid evolution of the fast wind. The tight constraints placed
    on the (nonsolar) abundances directly implicate the present-day
    central star --- hence, ultimately, the intershell region of the
    progenitor asymptotic giant branch star --- as the origin of the
    shocked plasma now emitting in X-rays.

  \end{abstract}


\section{Introduction}\label{sec1}
Planetary nebulae (PNe) are the last stages of evolution for
intermediate-mass stars (1-8 M$_{\odot}$). The central star that
generates a PN terminates its evolution as a cool asymptotic giant
branch (AGB) star by ejecting its outer envelope. The UV radiation of
the newly exposed hot core --- a future white dwarf (WD) --- then
illuminates and ionizes the ejected envelope.  At about the same time,
a ``hot bubble'' may be produced by collisions between the slowly
expanding, ambient AGB gas and the newly-initiated fast central star
wind (Kwok et al. 1978). Theories describing such wind
  interactions within PNe predict that the plasma temperature within
  the hot bubble should be high enough for the generation of soft
  ($\stackrel{<}{\sim}$ 1 keV) X-ray emission, and that the dimensions
  of the hot bubble X-ray source should be smaller than that of the
  optically-emitting, ionized nebula (e.g., Zhekov \& Perinotto 1996;
  Soker \& Kastner 2003; and references therein). ROSAT X-ray
  observations of PNe appeared to offered early support for these
  predictions (Kreysing et al. 1992; Guerrero et al. 2000; however see
  also Chu et al. 1993).

The advent of Chandra and XMM-Newton has
  provided far more convincing evidence for the presence of wind-blown
  hot bubbles within PNe (Kastner et al.\ 2008 and refs.\ therein;
  hereafter, K08).
However, certain puzzling aspects of the wind-collision-generated hot
bubbles within PNe remain to be explained. In particular, contrary to
the expectations of simple wind-collision models, the temperature of
the X-ray emitting hot bubbles in PNe does not appear to depend on the
 present central star wind velocity (K08). Furthermore, the
optical and X-ray 
emitting regions of the same object can display sharp differences in
abundances (e.g., Maness et al. 2003 and refs.\ therein). These
observations raise fundamental questions, e.g.: what heating and
cooling mechanisms govern the temperature of the X-ray emitting
plasma? Does the X-ray emission emanate primarily from the former AGB
star wind, the present central star wind, or some mixture of the two?

BD+30$^{\circ}$3639 (``Campbell's Star'') is a young planetary nebula
with a carbon Wolf-Rayet([WC]-type) central star; it has been studied
at a wide variety of wavelengths (e.g., Li et al.\ 2002 and references
therein). The nearby (distance $\sim1.2$ kpc; Li et al. 2002)
BD+30$^{\circ}$3639 has a young dynamical age ${\sim}$ (700 yr;
Leuenhagen et al.\ 1996), and its fast wind speed is ${\sim}$ 700 km
s$^{-1}$ (Leuenhagen et al. 1996; Marcolino et al.\ 2007). It is an
excellent target for X-ray observations, due to its unusually large
soft X-ray flux at earth (F$_{X}$ ${\sim}$ 4${\times}$10$^{-13}$ erg
cm$^{-2}$ s$^{-1}$, 0.1-2.0 keV; Kreysing et al. 1992; Arnaud et
al. 1996). Kreysing et al.\ (1992) first detected X-rays from
BD+30$^{\circ}$3639 with ROSAT, and estimated the hydrogen column
density toward and plasma temperature within the X-ray nebula ($N_{H}$
${\sim}$ 1.4 ${\times}$ 10$^{21}$ cm$^{-2}$ and $T_{x}$ ${\sim}$ 2.5
${\times}$ 10$^{6}$ K, respectively). The latter result ruled
  out, e.g., a hot companion to the $\sim30$ kK central star as the
  X-ray source and suggested the presence of a wind-shock-generated
  hot bubble within BD+30$^{\circ}$3639.

  Using the Chandra Advanced CCD Imaging Spectrometer (ACIS), Kastner
  et al. (2000) demonstrated that X-ray emission from
  BD+30$^{\circ}$3639 was spatially extended, and that the X-ray
  emitting gas is fully confined within the ${\sim}5''$ diameter
  elliptical ring of photoionized gas seen in optical and IR images.
  While these results supported the existence of a
    ``classical'' hot bubble within BD+30$^{\circ}$3639, one could
  not rule out other possibilities, such as jets resulting from binary
  interactions (Bachiller et al. 2000; Kastner et al. 2001, 2002;
  Soker \& Kastner 2003; Akashi et al. 2008).  Whether the X-ray
  emitting gas comes from jets or a fast spherical wind, there remains
  the question of the role, if any, of heat conduction between the
  X-ray emitting gas and the visible shell (Soker 1994; Zhekov \&
  Perinotto 1996; Zhekov \& Myasnikov 1998, 2000), and/or mixing 
  of the two media (Chu et al. 1997) enhanced by instabilities
  (Steffen et al. 2005, 2008; Stute \& Sahai 2006; Sch\"{o}nberner et al.\
  2006), in moderating the X-ray temperature to levels well below that
  expected from collisions between the present-day 700 km s$^{-1}$
  stellar wind and the previously ejected AGB star envelope.

  The source of the X-ray-emitting gas in BD+30$^{\circ}$3639 also
  remains to be determined, even though it has been a favorite subject
  of X-ray CCD spectroscopy.  Arnaud et al. (1996) obtained estimates of
  X-ray plasma abundances using ASCA CCD imaging spectrometer data,
  finding C, N and Ne to be significantly overabundant and Fe to be
  significantly depleted. Similar results were obtained via analysis
  of Chandra and Suzaku CCD imaging spectroscopy (Kastner et al.\
  2000; Maness et al.\ 2003; Murashima et al.\ 2006). However, these
  results are somewhat at odds with those obtained from optical/IR
  wavelengths that show, e.g., depleted Ne in the bright shell of the
  PN. While the presence of an enhanced Ne abundance in the
  X-ray-emitting plasma seems reasonably secure, the degree of Ne
  overabundance as well as other abundance
    anomalies --- such as highly enhanced C and highly depleted Fe,
    also inferred on the basis of X-ray CCD spectral modeling ---
  remain quite uncertain. Indeed, Georgiev et al. (2006) argued that
  X-ray CCD spectra cannot provide definitive constraints on the
  plasma abundances in PNe.

  To make progress on these and other problems concerning the nature
  and origin of the X-ray-emitting plasma within PNs requires X-ray
  observations at high spectral resolution, from which we
  unambiguously infer, with improved precision, the temperature and
  composition of the X-ray emitting plasma. With this motivation, we
  obtained a deep observation of BD+30$^{\circ}$3639 using
  Chandra's Low Energy Transmission Gratings spectrometer in
  combination with its Advanced CCD Imaging Spectrometer
  (LETG/ACIS-S). We selected LETG/ACIS-S (as opposed to HETG/ACIS-S or
  LETG/HRC-S) for Chandra gratings observations of BD+30$^{\circ}$3639
  on the basis of the superior background rejection, soft X-ray
  sensitivity, and order-sorting capabilities of this
  configuration. In addition to the dispersed spectrum of
  BD+30$^{\circ}$3639 (for which preliminary results were presented in
  Kastner et al.\ 2006), the LETG/ACIS-S observations produced a
  highly sensitive, undispersed 0th-order image of the PN. In this
  paper, we present a comprehensive analysis of the dispersed
  LETG/ACIS-S spectrum of BD+30$^{\circ}$3639. In a subsequent paper
  (Yu et al.\ 2008, in prep.) we present a spatial/spectral analysis
  of the LETG/ACIS-S zeroth-order image of BD+30$^{\circ}$3639 and
  compare this image with the direct Chandra/ACIS-S3 
    image obtained in Cycle 1 observations.

\section{Observations and Data Reduction}\label{sec2}

We obtained observations of BD+30$^{\circ}$3639 totaling 300 ks
exposure time with LETG/ACIS-S in 2006 February (85.4 ks), March (61.8
ks) and December (150 ks). The last observation was obtained in three
consecutive blocks at the same roll angle. The event data were
  subject to standard pipeline processing (using Chandra X-ray Center
    pipeline versions 7.6.7 for the 2006 February and March data and
    7.6.9 for the 2006 December data). In
Table~\ref{tbl-1}, we list exposure times and total photon counts
within the source and background regions. To determine total
counts in the 0th-order images, we selected a $10''$ circular region
centered on the source and a surrounding annular region (with inner
and outer radii of 13$''$ and 20$''$, respectively) for background.
The total 1st-order counts were determined directly from the extracted
source and background spectra (Sec.~\ref{sec3}). We also
extracted light curves from the source regions of the 2000 image and
the 0th-order images obtained in 2006; no measurable variability was
found, as expected given the diffuse nature of the source.

\begin{figure}
\epsscale{.80}
\centering
\includegraphics [scale=0.425, angle=90]{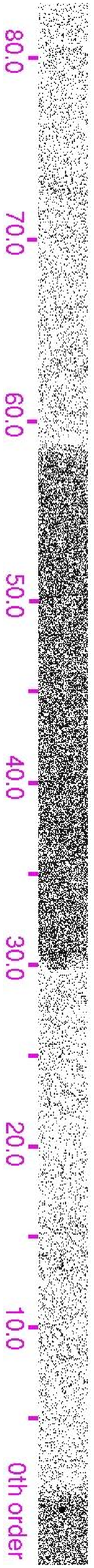}
\vskip 0.3in
\includegraphics [scale=0.500, angle=90]{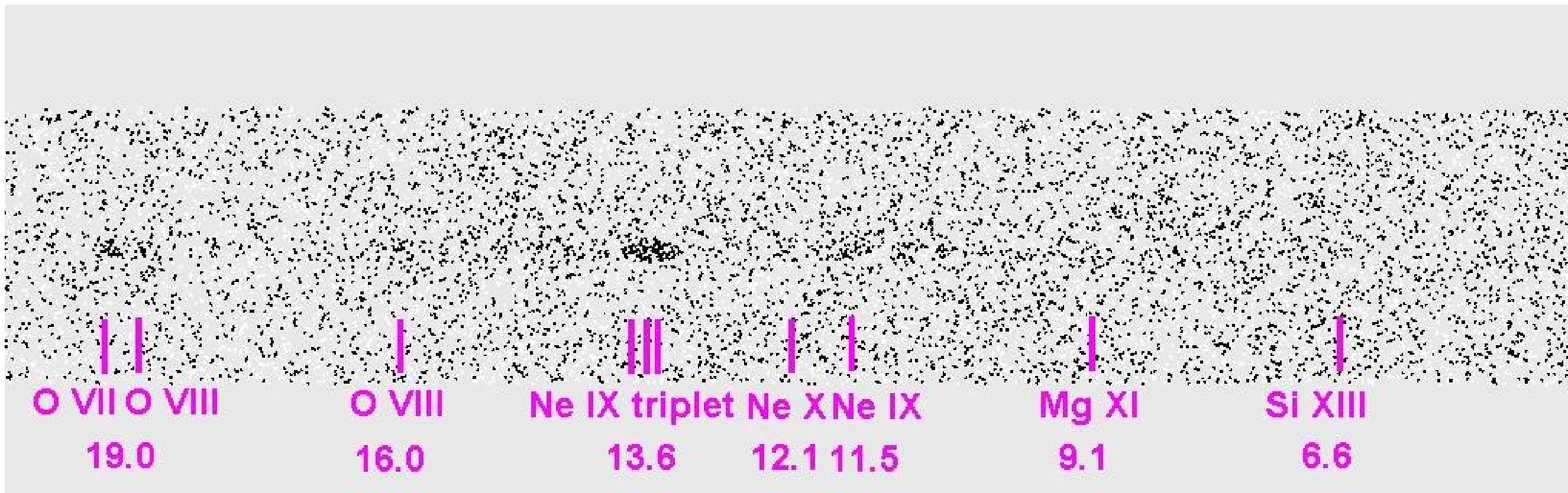}
\vskip 0.1in
\includegraphics [scale=0.500, angle=90]{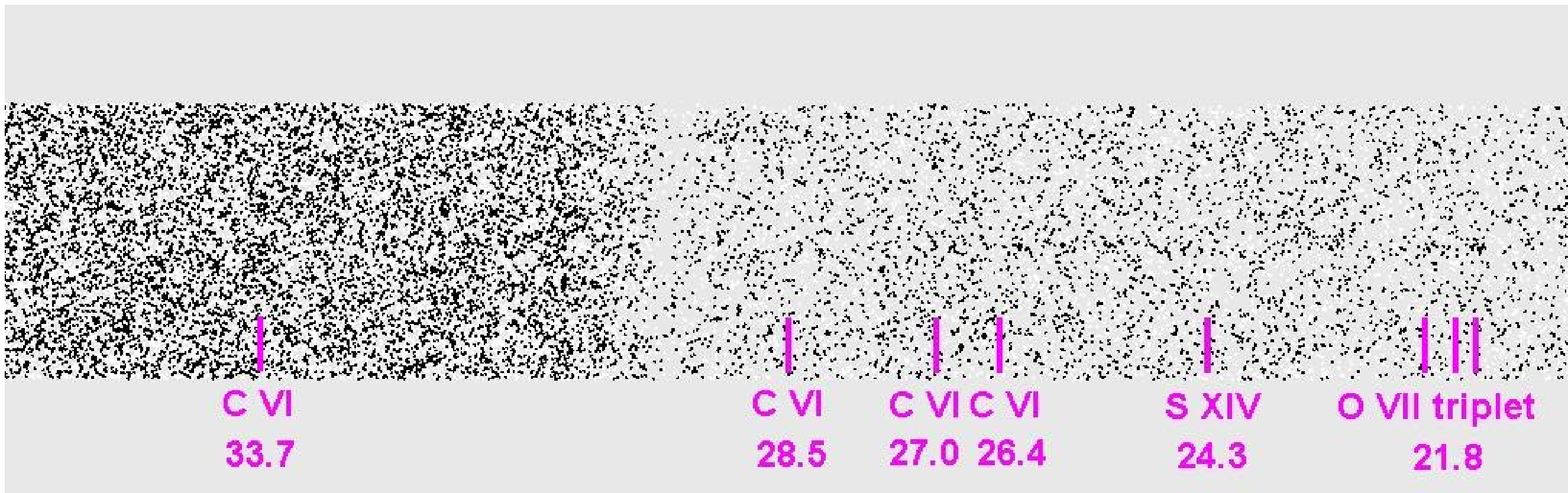}
\caption{Dispersed spectral images of BD +30$^\circ$3639 for negative
  LETG orders, with wavelength scales (in \AA) overlaid. Top: full range of
  negative orders. Middle and bottom: closeups of dispersed spectral
  images over the wavelength ranges 5-20 \AA\ and 20-35 \AA,
  respectively.  \label{dispNegaImage}}
\end{figure}

\begin{figure}
\epsscale{.80}
\centering
\includegraphics [scale=0.425, angle=90]{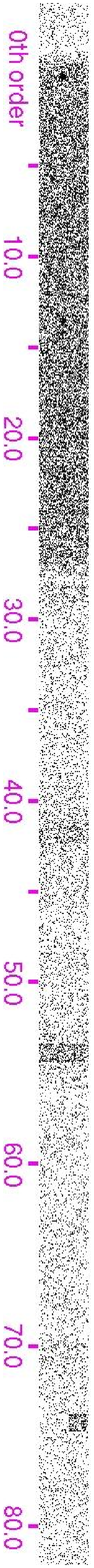}
\vskip 0.3in
\includegraphics [scale=0.500, angle=90]{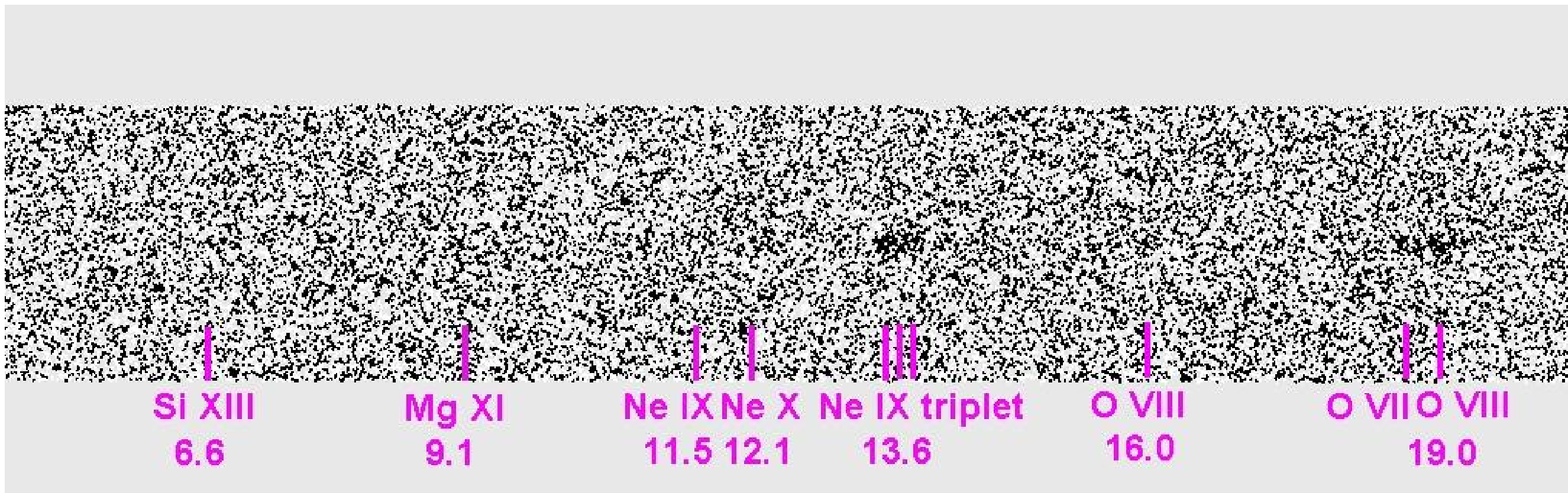}
\vskip 0.1in
\includegraphics [scale=0.500, angle=90]{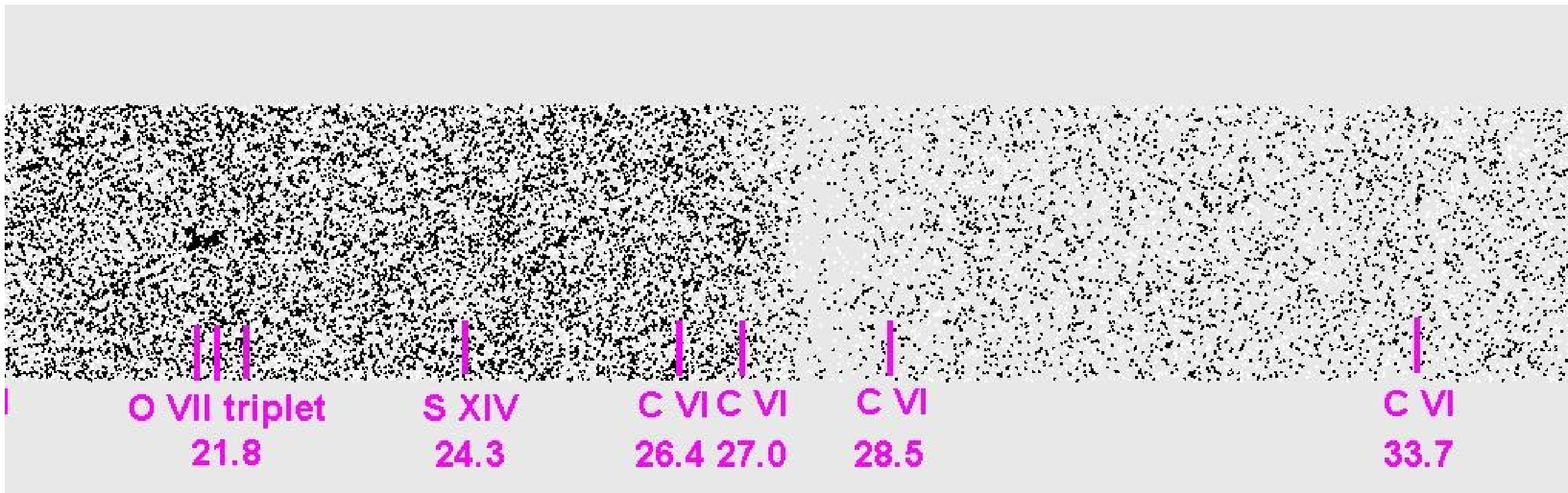}
\caption{As in Fig.\ 1, for positive LETG orders.  \label{dispPosiImage}}
\end{figure}

Because the first two observations in 2006 (ObsIDs 5409 \& 7278) were
obtained at different spacecraft roll angles and aimpoints with
respect to each other and the last three (Dec.) 2006 observations, we
cannot generate a merged spectral image for the full 300 ks
exposure. However, the second (Dec.\ 2006) half of the 300 ks exposure
(ObsIDs 5410, 8495 \& 8498) was obtained at constant roll angle, so we
generated a single, 150-ks-exposure dispersed spectral image from
these data by merging the three Level 2 event (\verb+evt2+) files. The
full spectral images obtained from the resulting combined
\verb+evt2+ files are shown in Figures~\ref{dispNegaImage} and
~\ref{dispPosiImage}. These dispersed images demonstrate that the
LETG/ACIS-S X-ray counts spectrum of BD+30$^{\circ}$3639 is dominated
by emission lines of highly ionized (He-like and H-like) ions of
oxygen and neon.

For analysis of the dispersed spectrum, further processing involved
removal of artifacts on ACIS-S4 (using the CIAO tool ``destreak'') and
applying updated calibrations. Standard gratings point source
  spectral extraction threads available in
  CIAO\footnote{http://cxc.harvard.edu/ciao/threads/spectra\_letgacis/}
were used to generate spectrum pulse height amplitude (PHA) files and
corresponding redistribution matrix files (RMFs) and auxiliary
response files (ARFs). These threads implicitly ignore the
  (non-negligible, $\sim5''$) spatial extension of the
  BD+30$^{\circ}$3639 X-ray source, resulting in artifically
  ``broadened'' emission line features (we discuss this effect further
  in Secs.\ 3.1 and 3.2, respectively). We extracted positive ($m=
+1$) and negative ($m= -1$) first-order LETG spectra for each of the
five observations (Table~\ref{tbl-1}). To enhance the signal to noise
ratio, each LETG spectrum was rebinned by a factor 8, resulting in a
wavelength dispersion of 0.1 \AA/bin. We then merged the resulting
spectra into a single spectrum. In parallel, the corresponding ARFs
were averaged and weighted by the relative exposure times, then merged
into a single ARF. The various RMFs obtained from the individual
observations for a given order are identical to within the calibration
uncertainties. Hence, for the spectral analysis described below, we
used the same, representative RMF for each LETG dispersion arm.

\section{Analysis and Results}\label{sec3}

\begin{figure}
\epsscale{.80}
\centering
\includegraphics [scale=0.55, angle=270]{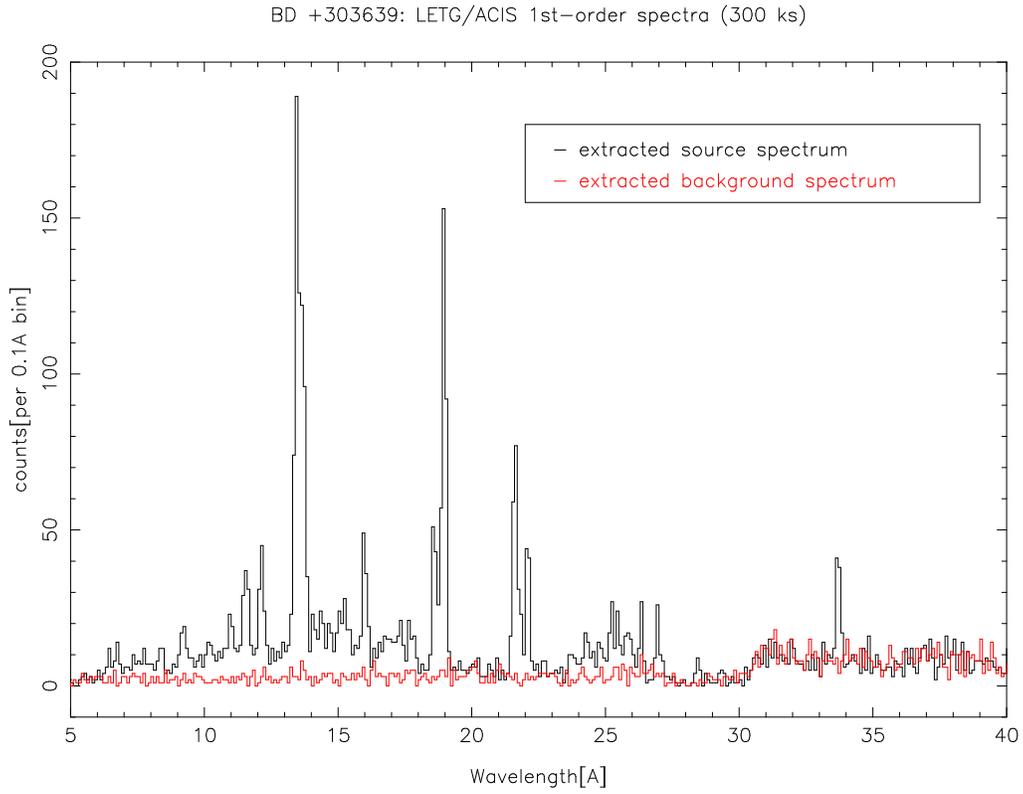}
\vskip 0.2in
\includegraphics [scale=0.55, angle=270]{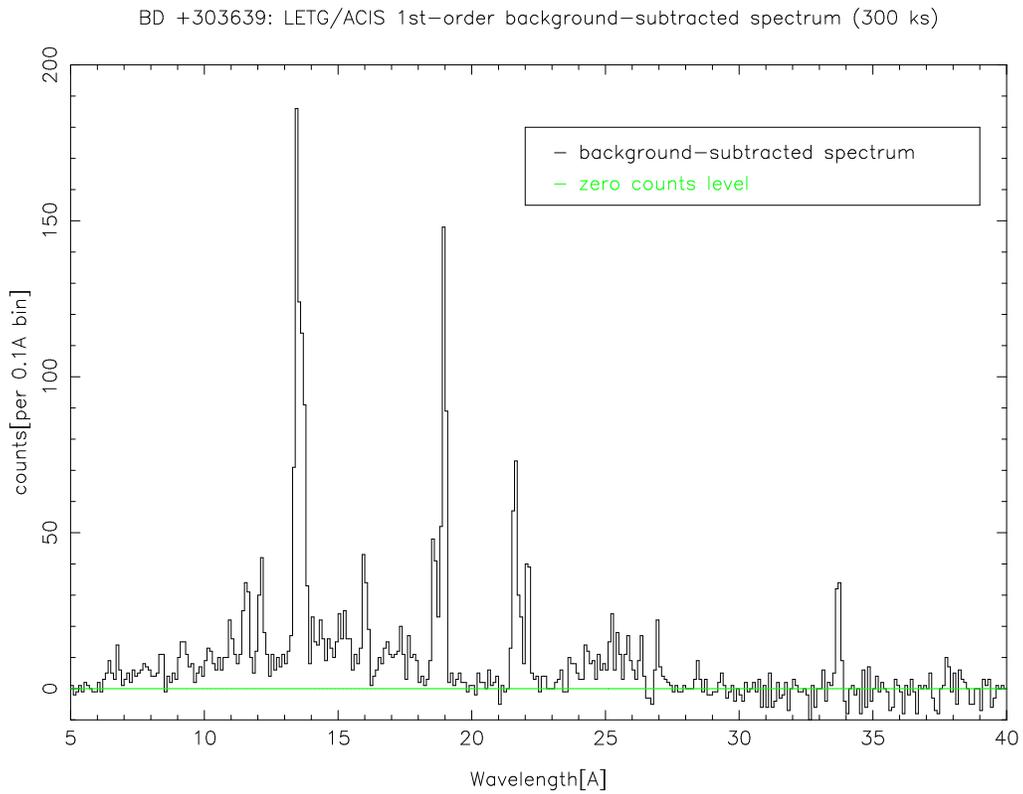}
\caption{(a) Combined positive and negative first order LETG/ACIS-S counts 
spectra of BD+30$^{\circ}$3639 (black) and background (red). (b)
Background-subtraction counts spectrum of
BD+30$^{\circ}$3639.\label{combFirstorder}}  
\end{figure}

The merged, 300 ks exposure time spectrum resulting from the LETG/ACIS
spectral image data reduction procedure described in \S 2 is displayed
in Fig.~\ref{combFirstorder}a. The spectrum displays strong emission
lines superposed on a weak continuum, with an abrupt rise in the
continuum at $\sim30$ \AA\ that is likely due to background events. To
account for this apparent residual background, we identified a region
devoid of bright X-ray sources, displaced 215 arcsec from the source
position along the detector y direction, and extracted first-order
background spectra at this position from each of the 5
observations. The corresponding RMFs and ARFs of background spectra
were also generated. We rebinned and merged these individual
background spectra into a single spectrum (seen superimposed on the
source spectrum in Fig.~\ref{combFirstorder}a).
Figure~\ref{combFirstorder}b shows the resulting combined, 1st-order,
background-subtracted spectrum of BD+30$^{\circ}$3639 in the
wavelength range of 5 - 40 \AA. The apparent ``continuum'' in the
region 30 - 40 {\AA} is effectively removed from this
background-subtracted spectrum.

The brightest lines in the background-subtracted spectrum of
BD+30$^{\circ}$3639 are due to highly-ionized (H- and He-like)
  species of C, O, and Ne; 
the resonance lines of H-like O {\sc viii}
(${\lambda}$ 18.97) and C {\sc vi} (${\lambda}$ 33.6) and the He-like
triplet line complexes of Ne {\sc ix} (${\lambda}$ 13.45, 13.55, 13.7) and O
{\sc vii} (${\lambda}$ 21.60, 21.80, 22.10) are especially
  prominent. Other H-like resonance lines, 
such as N {\sc vii} (${\lambda}$ 24.78) and lines of highly ionized Fe, are
weak or absent. Continuum emission is evident over the range 6--18
\AA. The excess emission near 25 {\AA} can likely be attributed to enhanced  
high-order lines of C {\sc vi} within its H-like recombination
  line spectrum (Nordon et al.\ 2008, in prep.).

\subsection{Line identifications, fluxes, and source angular sizes}\label{sec3.1} 

We measured line fluxes with the Interactive Spectral Interpretation
System (ISIS\footnote{http://space.mit.edu/CXC/ISIS/}; Houck \&
Denicola 2000). Given the relatively symmetric appearances of the
profiles of prominent emission lines, we used a fit function
consisting of a constant local continuum (polynomial) plus one or more
Gaussian functions. In fitting the strong He-like triplet lines (Ne
{\sc ix} and O {\sc vii}), the ratios of line strengths and widths of
the triplet components were fixed using the values given by the
Chandra atomic database (ATOMDB ver. 1.3) for the case of
  low-density plasma ($n_e << 10^{10}$ cm$^{-3}$). Thus, the free
parameters were the intensities of triplets, the line-center
wavelength and FWHM of one of the triplets, and the coefficients of
the polynomial representing the local continuum.

For those emission lines that could be measured with acceptable
statistics, Table~\ref{tbl-2} lists the line identifications, fluxes,
and widths. The line-fitting procedure thereby confirms the
  identification of at least 15 lines and line complexes in the LETG
  spectrum of BD+30$^\circ$3639, ranging from the very strong C {\sc
    vi} L$\alpha$ line to weak Mg {\sc xii} and Si {\sc xiii}
  lines. The Table also lists upper limits on the fluxes of the lines
  of other ions that are of similar ionization potential to the
  well-detected species such as Ne {\sc ix}, Ne {\sc x}, O {\sc vii}, O
  {\sc viii}, and C {\sc vi}.  Lines of these important species ---
  e.g., Fe {\sc xvii} and N {\sc vii} --- should be prominent 
  in the spectrum of a solar-abundance plasma at the
  approximate temperature implied by the well-detected lines ($T_x
  \sim 2\times10^6$ K).

  Since the X-ray line widths reflect the extended nature of the
  source rather than, e.g., plasma turbulence (see below) or
  kinematics, the width
  measurements for well detected lines ($\Delta \lambda$) are
  expressed in terms of
  the corresponding angular FWHM in arcsec, assuming a dispersion for
  LETG of 18.02 arcsec {\AA}$^{-1}$ (Dewey, 2002).  The resulting angular
  FWHMs are consistent with each other and with the  ($\sim5''$)  source
  angular extent in the zeroth-order image, within the
  uncertainties. Furthermore there are
  no discernable systematic redshifts or blueshifts measured for the
  emission line centers. Hence, for purposes of the plasma modeling
  described here, all of the emission lines can be considered to arise from
  the same region within the nebula.

\subsection{Global spectral fitting}\label{sec3.2}

Global X-ray spectral model fitting of lines and underlying continuum
is necessary to simultaneously constrain relative plasma elemental
abundances and temperatures. In adopting this approach to fit the LETG
spectrum of BD+30$^{\circ}$3639, we selected the {\it ``Cash''} method
(Cash, 1979) as the fit statistic to treat low count data.  For the
model fitting, we specified the displaced background spectrum (see \S
3.1) along with the merged source spectrum. To investigate plasma
physical conditions, we used ISIS to construct Astrophysical Plasma
Emission Database (APED; Smith et al. 2001) models, varying plasma
 metal abundances such as Fe, Ne, O, C, N and Mg, and leaving all
other abundances, including H and He, fixed at solar (Anders \&
Grevesse 1989). The latter assumption is, of course, unlikely to
  be valid if the X-ray-emitting plasma is dominated by the
  H-depleted, He-enriched wind characteristic of the present-day
  central star (e.g., Marcolino et al.\ 2007) --- a hypothesis that is
  indeed supported by our modeling results (see \S\S 3.2.2, 4.1).  We
did not consider higher Lyman series C {\sc vi} lines that lie in the
region $\sim$ 25--29 {\AA} in the spectral model because APED only
includes the C {\sc vi} Lyman series up to the $\delta$ line. Analysis
of these higher-energy C {\sc vi} transitions is described in Nordon
et al.\ 2008 (in prep.).

\subsubsection{Isothermal vs.\ two-component APED models}

To reproduce the merged LETG 1st-order spectrum of BD +30$^\circ$3639,
we attempted fits with both single-component and two-component APED
plasma models. Note that the spectral resolving power of LETG for
  a point source ranges from $E/{\bigtriangleup}E \sim$130 at 5 {\AA}
  to $E/{\bigtriangleup}E \sim$1000 at 40 {\AA} (Dewey 2002
  ). The spatial extension of BD+30$^{\circ}$3639
  (FWHM ${\sim}$3 arcsec; Table 2), combined with use of ACIS-S rather
  than HRC as the detector, then degrades the effective resolving
  power of LETG by a factor $\sim3$ (Dewey 2002). Source spatial
  extent --- rather than, e.g., gas turbulence or thermal broadening
  --- therefore determines the line widths. To emulate this artificial
  line ``broadening'', we used the turbulent velocity parameter
($V_{turb}$) available in the APED model. We find that two turbulent
velocity components --- with best-fit parameter values of $\sim1700$
km s$^{-1}$ and $\sim900$ km s$^{-1}$ for the short- and
long-wavelength spectral regions, respectively, consistent with the
mean FWHM measured for the lines --- are sufficient to reproduce the
``broadening'' caused by the extended X-ray source within BD
+30$^\circ$3639. We emphasize that these results do not actually
  represent measurements of gas turbulence; rather, like emission line
  FWHM (Table 2), the best-fit $V_{turb}$ values serve as an
  indication of source spatial extent.

Each of the models was assumed to undergo absorption due to
intervening neutral material characterized in terms of the column
density of neutral H, $N_H$,
  using the standard, solar-abundance (\verb+wabs+) model (Morrison \&
  McCammon 1983). However note that, given
  the close correspondence
  between the visual extinction and X-ray surface brightness
  distributions of BD+30$^\circ$3639 (Kastner et al.\ 2002), the
  absorption is in fact best
  attributed to the nebula itself rather than
  intervening interstellar medium. The X-ray-absorbing material may be the
  extended molecular envelope of BD+30$^\circ$3639, pockets of cold,
  dense gas embedded in the ionized nebula, or some combination of
  these contributions; as a result, the composition of the absorbing
  material may differ significantly from that assumed in the
  \verb+wabs+ model. The implications for the results for $N_H$, as
  well as for model plasma abundances, are discussed below.

\begin{figure}
\epsscale{.80}
\centering
\includegraphics [scale=0.55, angle=270]{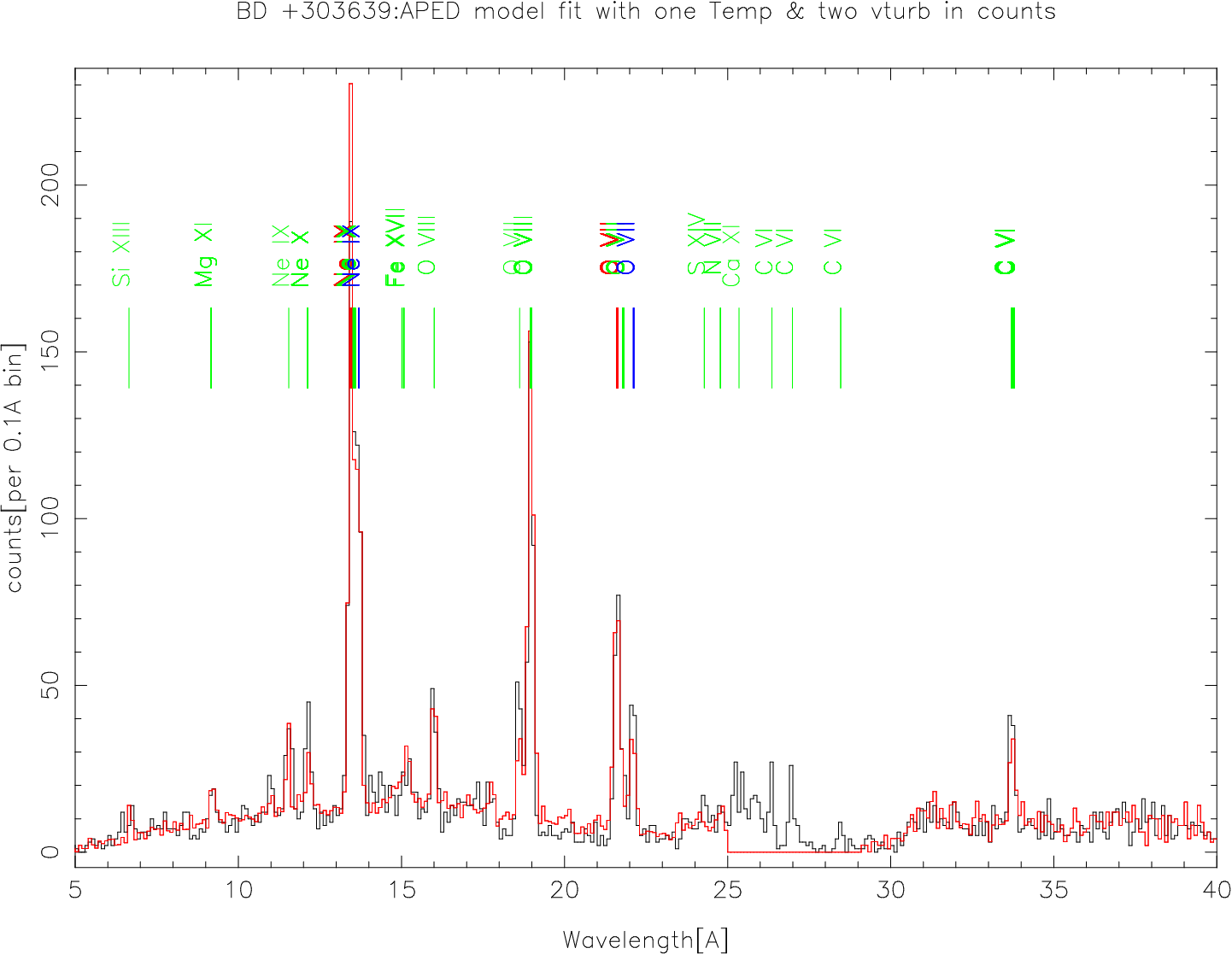}
\vskip 0.2in
\includegraphics [scale=0.55, angle=270]{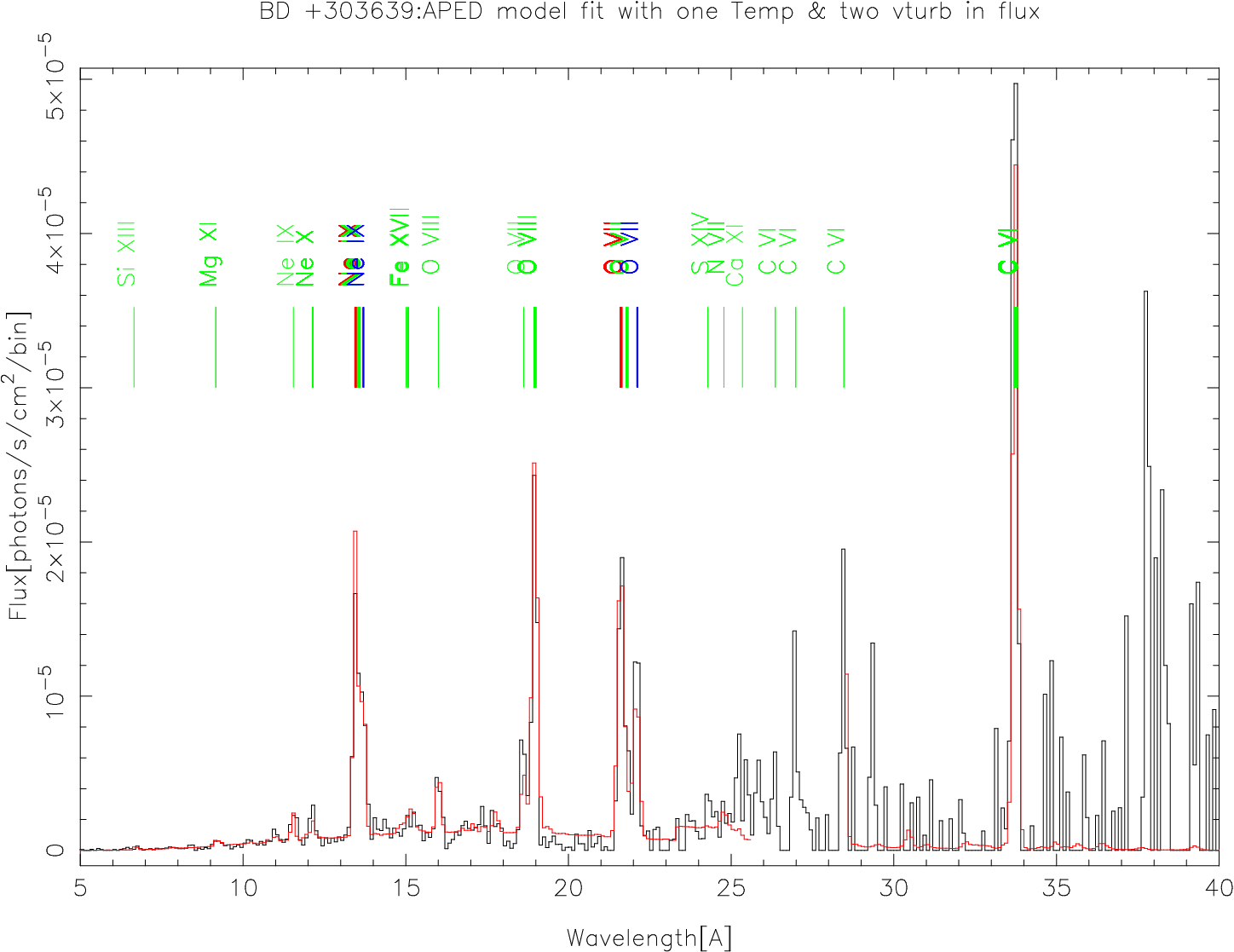}
\caption{(a) Combined positive and negative first-order LETG/ACIS-S
  counts spectrum of BD+30$^{\circ}$3639 (including background)
  overlaid with the best-fit single-component APED model. (b) Flux-calibrated
  first-order LETG/ACIS-S spectrum, overlaid with the same model. In
  each panel, black shows the source spectrum and red indicates the
  model. \label{One_2Vturb_model}}
\end{figure}

Figure~\ref{One_2Vturb_model}a shows the merged, 300 ks exposure, 1st-order
LETG/ACIS-S counts spectrum of BD+30$^{\circ}$3639 overlaid with the
single-component model spectrum obtained from the APED. The comparison
of the flux-calibrated spectrum with the model makes apparent the
strength of C {\sc vi} (${\lambda}$ 33.6) relative to the other strong lines
(Ne {\sc ix}, O {\sc vii} and O {\sc viii}; 
Figure ~\ref{One_2Vturb_model}b). Other features
apparent in the long wavelength (${\lambda}$ $>$ 30) region of the
flux-calibrated spectrum are likely artifacts of poor
  photon counting statistics combined with 
the very low net effective area of LETG/ACIS in this region.
However, we find this single-component APED model is
insufficient to adequately fit the spectrum. In particular, while the
O {\sc vii} to O {\sc viii} line ratios are reasonably well
reproduced, the model cannot simultaneously fit the Ne {\sc ix} and Ne {\sc x} 
lines in the 12 \AA\ region. 

In contrast, the two-component APED model well reproduces the
intensities of all strong emission lines and better matches the
6--18 \AA\ continuum (Fig.~\ref{Two_comp_model}). The high-temperature 
boundary is constrained in part by the nondetection 
of the Mg {\sc xii} line at 8.42 {\AA} contrasted with 
the weak but clear Mg {\sc xi} line at 9.12 {\AA}. The 
low-temperature boundary is harder to constrain, as it
relies in part on the relative intensities of the (somewhat noisy) C
{\sc vi} lines at 28.46 {\AA} and 33.73 {\AA}. Since the plasma
temperatures are therefore mainly governed by two indicators --- i.e.,
the line ratios of the H-like to He-like O and Ne --- introducing more
than 2 temperature components into the model would make the fit result
degenerate. 

\subsubsection{Best-fit parameters: results and confidence ranges}

The best-fit temperature of the single-component APED model indicates
that, under the isothermal plasma approximation, the characteristic
plasma temperature lies in the range 2.2--2.4$\times10^6$ K. It is
therefore not surprising that the values of $T_X$ obtained from the
two-temperature-component model ($T_1 = 2.9\times10^6$ K and $T_2 =
1.7\times10^6$ K) brace this range.  Table 3 demonstrates that the two
preceding models also yield consistent results where plasma abundances
are concerned but that the two-component model, in addition to
providing a superior fit to the LETG spectrum, yields these abundance
results to greater precision. 

  We reiterate that the best-fit absolute abundances listed in
  Table 3 (as number ratios relative to solar) were obtained under the
  assumption of solar H and He abundance (Anders \& Grevesse 1989) and
  are therefore subject to large but unknown systematic
  uncertainties. Specifically, no diagnostics of H and He abundances
  are available in the X-ray regime, yet it is likely that the
  X-ray-emitting plasma is dominated by present-day (H-depleted,
  He-enriched) [WC] stellar wind material (\S 4.1).  In this case of a
  strongly H-depleted stellar wind, the abundance normalization to H
  becomes irrelevant. One could, however, re-normalize the Table 3 results
  to the solar He abundance under the assumption that ratio of the C to He
  abundances is identical to the [WC] stellar wind value, i.e.,
  C/He $\sim$ 0.4 by number (Marcolino et al. 2007), or $\sim$108 times
  the solar ratio.  The values listed in Table 3 could then be
  increased by a factor 108/28.3 = 3.8, where 28.3 is the C abundance obtained
  from the two-component model fit, 
  so as to yield elemental 
  abundances relative to He in
  solar units. The results for emission measure would then also have
  to be redefined to refer to the number density of He nuclei (rather
  than H). In the context of the APEC model, this is equivalent to
  reducing the model normalization by the same factor as that of the
  increase in elemental abundances.

Given the foregoing problems in attempts to determine absolute
  abundances, all subsequent analysis in this paper is based on the
  modeling results for abundances as number ratios relative to O. In
  this regard, the results listed in Table 3 then indicate
  that C is very overabundant (C/O $\sim30$, relative to the solar
  ratio), Ne is overabundant (Ne/O $\sim3.8$), and both Fe and N
  likely are depleted (Fe/O $\sim0.2$ and N/O $\sim0.4$).

Because the LETG response matrix is nearly diagonal, the observed
(absorbed) source X-ray flux deduced from the model fitting is similar
 for the two models ($F_X= 4.4\times10^{-13}$ ergs cm$^{-2}$
s$^{-1}$). The inferred intrinsic (unabsorbed) X-ray source luminosity
is then constrained by the model fitting to lie in the range
$7.4\times10^{32}$ erg s$^{-1}$ to $8.6\times10^{32}$ erg s$^{-1}$. The 
low-$T$ and high-$T$ components of the latter
model account for $\sim$25\% and $\sim$75\% of the total source
luminosity, respectively.

\begin{figure}
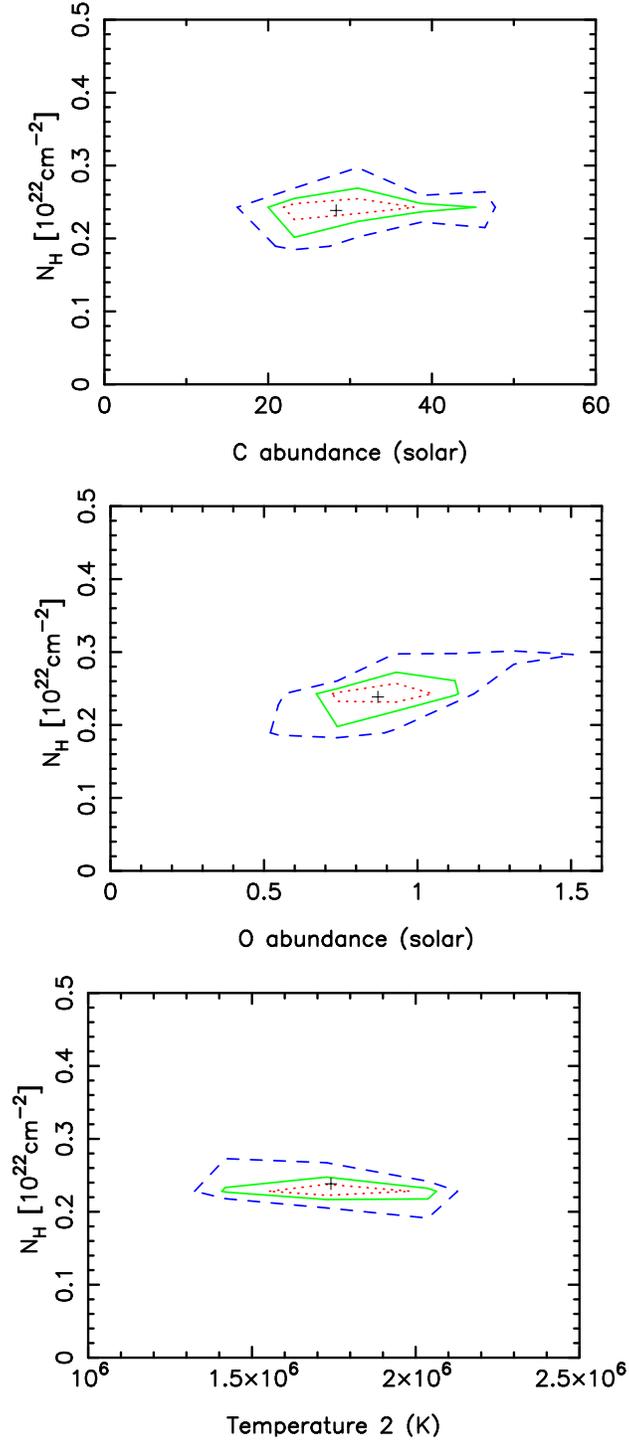

\epsscale{.80}
\centering
\includegraphics [scale=0.35, angle=270]{f6a.eps}
\vskip 0.1in
\includegraphics [scale=0.35, angle=270]{f6b.eps}
\vskip 0.1in
\includegraphics [scale=0.35, angle=270]{f6c.eps}
\caption{Plots of best-fit confidence contours (68\%, dotted; 90\%, solid; and
  99\%, dashed), as obtained from the two-component
 APED model fitting, for the column density parameter $N_{H}$ vs. the C
  abundance parameter (top), O abundance parameter (middle),
 and the lower of the two temperatures (bottom). 
\label{confidence_contour_vs_nH}}
\end{figure}

\begin{figure}
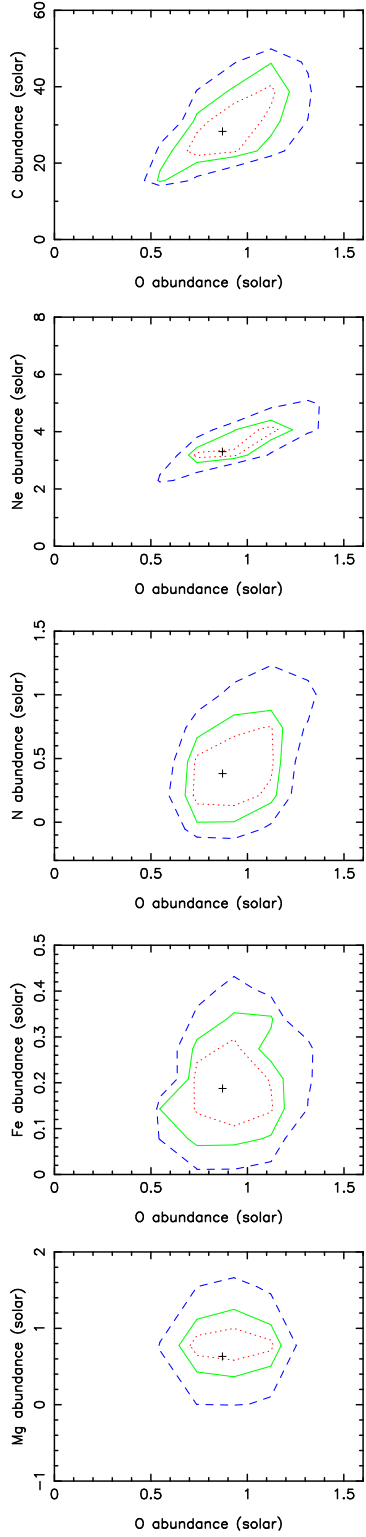

\epsscale{.80}
\centering
\includegraphics [scale=0.22, angle=270]{f7a.eps}
\vskip 0.1in
\includegraphics [scale=0.22, angle=270]{f7b.eps}
\vskip 0.1in
\includegraphics [scale=0.22, angle=270]{f7c.eps}
\vskip 0.1in
\includegraphics [scale=0.22, angle=270]{f7d.eps}
\vskip 0.1in
\includegraphics [scale=0.22, angle=270]{f7e.eps}
\vskip 0.1in
\caption{As in Fig.~\ref{confidence_contour_vs_nH} for the C, Ne, 
N, Fe, and Mg abundance parameters (top to bottom, respectively) vs. the 
O abundance parameter.
 \label{confidence_contour_vs_O}}
\end{figure}

\begin{figure}
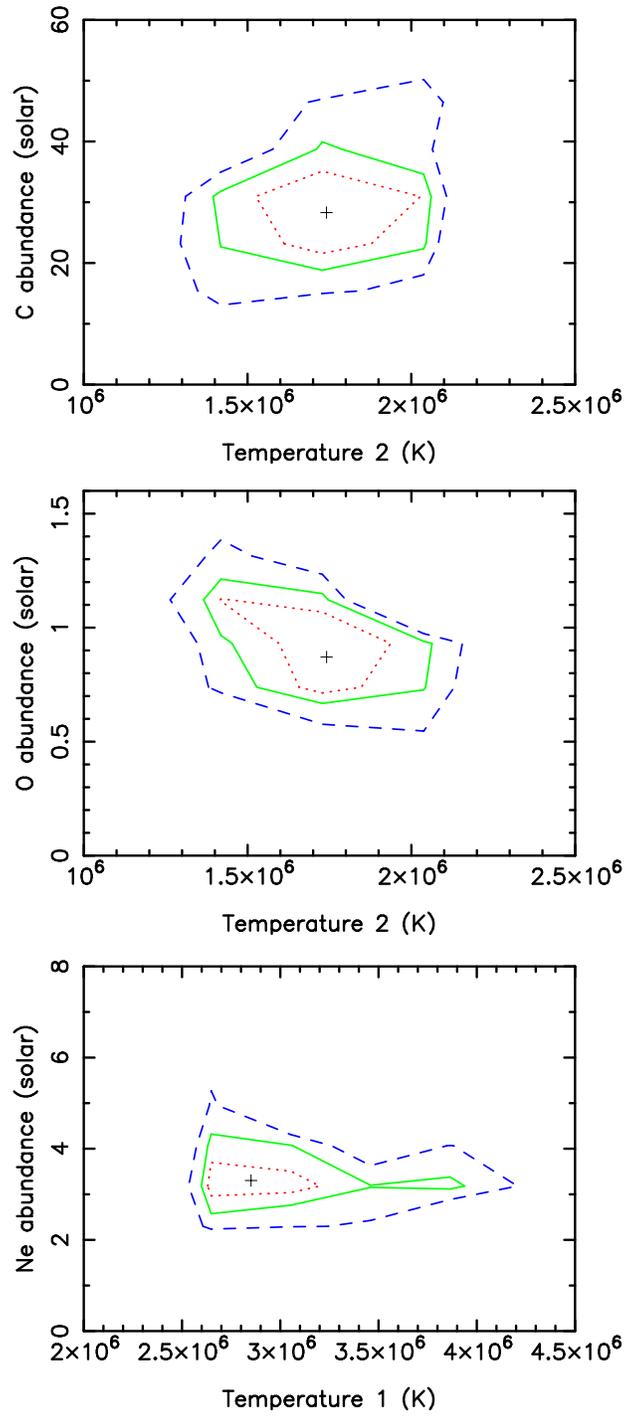

\epsscale{.80}
\centering
\includegraphics [scale=0.35, angle=270]{f8a.eps}
\vskip 0.1in
\includegraphics [scale=0.35, angle=270]{f8b.eps}
\vskip 0.1in
\includegraphics [scale=0.35, angle=270]{f8c.eps}
\caption{As in Fig.~\ref{confidence_contour_vs_nH} for C and O
  abundances vs.\ the lower of
  the two temperatures (top and middle panels) and the Ne abundance
  vs. the higher of the two temperatures (bottom). 
 \label{confidence_contour_vs_T}}
\end{figure}

In Figs.~\ref{confidence_contour_vs_nH}-\ref{confidence_contour_vs_T}
we present plots of confidence contours obtained from the
two-component APED model for various parameter combinations.
Figure~\ref{confidence_contour_vs_nH} demonstrates that N$_H$, which
is constrained mainly by the relative strengths of metal lines in the
softer part of spectrum, is (in principle) very well determined, and
that this parameter is essentially insensitive to the plasma
temperature and C and O abundances obtained from the model fitting.
Given a standard (ISM) gas-to-dust (hence $N_H/A_V$) ratio, the
best-fit value of $N_H = 2.4\times10^{21}$ cm$^{-2}$ obtained from the
two-component model is roughly consistent with the typical visual
(dust) extinction measured toward the regions of the nebula from which
X-rays are detected ($A_V \sim$ 1--2; Kastner et al. 2002). Adopting
the (overly simplistic) assumption that the absorption arises in a
spherically symmetric envelope surrounding the X-ray-emitting plasma,
the best-fit value of $N_H$ in the two-component model implies a
neutral envelope mass of $\sim0.03 M_\odot$ --- in reasonable
agreement with available estimates (e.g., Bachiller et al.\ 1991). We
conclude that $N_H$ is unlikely to be overestimated due to the
assumption of solar abundances in the absorbing column (implicit via
use of the \verb+wabs+ model).

In addition, the confidence contour plots demonstrate that the C and
Ne abundance parameters are well correlated with O, indicating that
the C/O and Ne/O ratios (as well as other key abundance ratios, such
as Fe/O) are very well constrained by the LETG spectrum
(Fig.~\ref{confidence_contour_vs_O}) --- even though the individual
absolute abundances remain uncertain, due to the lack of constraints
on the plasma H abundance.  Specifically, we find C/O ${\sim}$15--45,
Ne/O ${\sim}$3.3--5, N/O ${\sim}$0--1.0, and Fe/O ${\sim}$0.1--0.4,
relative to the solar ratios. The existence of a large C overabundance
(relative to solar) is supported by the tight constraints placed on
N$_{H}$ and the lack of correlation between inferred C abundance and
N$_H$ (Fig.~\ref{confidence_contour_vs_nH}). That is, it is unlikely
that the inferred high C abundance can be attributed to an
overestimate of the intervening absorption due to neutral metals.
  On the other hand, as just discussed, it is also unlikely that the
  abundance of C has been vastly underestimated due to the assumption
  of standard (solar) metal abundances in the absorbing material.
  Fig.~\ref{confidence_contour_vs_T} futhermore makes clear that the
  inferred C, O, and Ne abundances are relatively insensitive to the
  best-fit plasma temperatures.

\section{Discussion}

\subsection{Plasma modeling: comparison with previous results}

  In Table 4 we compare previous results from X-ray CCD
  spectroscopy of BD+30$^\circ$3639 with those obtained from modeling
  its LETG/ACIS spectrum (Sec.\ 3.2).
  Whereas the value of $T_X$ obtained from the isothermal model
  (Table 3) is on the low side of the range found in
  the CCD-based work --- likely reflecting the spectral dominance of
  the longer-wavelength lines of C and O --- the temperatures
  determined from the two-component model fit brace the range of
  values previously determined from CCD spectra. Although the LETGS
  modeling definitively demonstrates that isothermal models are not 
  adequate to match the line spectrum in detail (\S 3.2.1), the
  comparison to previous work, as well as to the isothermal model
  explored here, suggests that single-component plasma models
  that are based on X-ray CCD spectra are capable of
  recovering the characteristic temperature (as opposed to the
  temperature extremes) of the superheated plasma in PNs.

  The LETG/ACIS spectral modeling confirms the Ne and C enrichment and
  Fe underabundance in the X-ray-emitting plasma of BD+30$^\circ$3639
  that were previously inferred from X-ray CCD spectra, albeit with
  much improved constraints on the degree of these abundance anomalies
  (relative to O). In particular, the LETG/ACIS spectral modeling ---
  while confirming that C is highly enriched in the X-ray emitting
  plasma -- definitively precludes a plasma C/O ratio larger
  than $\sim45$, relative to solar. Indeed, our lower limit on the C/O
  ratio in the diffuse X-ray emission (C/O $\sim15$) is much
    more consistent with the ratio at the central star, as derived
  from optical/UV spectroscopy (C/O $\sim12$; Leuenhagen et al.\ 1996;
  Marcolino et al.\ 2007), than with the ratio characteristic of
    the nebular gas (C/O $\sim1.6$; Pwa et al.\ 1986). 

  In contrast to the case of C, whose large overabundance is
    evident from the strength of the resonance line of C {\sc vi},
  there is no unambiguous 
  evidence for Fe emission lines in the entire LETG/ACIS spectrum of
  BD+30$^\circ$3639. Although a line is present at $\sim15.15$ \AA\
  that may be partly due to 15.014 \AA\ line emission from Fe {\sc
    xvii}, it is likely that the O {\sc viii} line at 15.1670 {\AA}
  contributes to (even dominates) the flux measured for the
    $\sim15.15$ \AA\ line. In
  addition, we do not clearly detect any Fe {\sc xvii} emission lines
  around 17\AA; these lines
  are expected to be bright at the relatively low plasma temperatures
  found here (Doron \& Behar 2002). Furthermore there is no evidence
  for lines of Fe {\sc xvi} (or Fe {\sc xviii}), such as might be
  expected if the temperature were too low (or too high) for efficient
  formation of Fe {\sc xvii} via ionization equlibrium. This lack of
  Fe lines results in a firm upper limit on Fe abundance of $\sim0.3$
  relative to solar (Table 3), consistent with the optical/UV
    results of Georgiev 
  et al. (2006) for the nebula and Marcolino et al. (2007) for the
  central star.

  In addition, the 24.8 \AA\ resonance line of N {\sc vii} is not
  detected, although the line flux is not as well constrained as that
  of Fe {\sc xvii} due to the low sensitivity of LETG/ACIS in the
  former wavelength regime. As a consequence of this nondetection, the
  N abundance obtained from the LETG/ACIS spectral modeling
  demonstrates that N is, if anything, underabundant relative to
  solar, contradicting several previous X-ray abundance studies. Our
  gratings-based result of an N underabundance, which is consistent with
  optical/UV results for the central star (Marcolino et al. 2007),
  indicates that modeling of previous (CCD) X-ray spectra has
  confused N with C. We also find that previous indications of
  depleted Mg obtained from CCD spectra are not supported by the LETG
  data.

  The absorbing column and (thus) intrinsic X-ray luminosity
  inferred from the LETG/ACIS modeling are on the high side of the
  range of values previously determined from X-ray CCD spectra
  (Table 4). However, the LETG spectral modeling has significantly
  decreased the uncertainty in the inferred value of N$_{H}$ and,
  hence, $L_X$. 

\subsection{Plasma abundances: constraints on the origin of the
  X-ray-emitting gas}

The similarity of the large overabundance of C and underabundance of
Fe we determine from the LETG/ACIS spectral modeling to these same
anomalies as determined for the central star (Marcolino et al.\ 2007
and references therein) traces the X-ray-emitting gas directly back to
the present-day central star. Indeed, the mass fractions one
  would deduce from Table 3 under the assumption that the C/He number
  ratio is identical to that of the central [WC] star --- 41.5\% He,
  49.7\% C, 4.8\% O, 3.3\% Ne, $<0.24$\% N, and $<0.2$\% Fe --- agree
  remarkably well with those found by Marcolino et al.\ (2007) for the
  [WC] stellar wind. Furthermore, our robust determination of a Ne
overabundance of $\sim3$ to $\sim6$ relative to solar is consistent
with the predictions of models describing H-deficient central stars of
PNe (Werner \& Herwig 2006).  Hence, as also pointed out by Murashima
et al.\ (2006) and Kastner et al.\ (2006), the nonsolar composition of
the X-ray-emitting plasma in BD+30$^{\circ}$3639 appears to be the
direct result of nucleosynthesis processes in the precursor AGB star
(e.g., Herwig 2005 and references therein).

These results indicate that the shocked plasma now seen in X-rays
  originated deep within the AGB star, in the ``intershell'' region;
He-shell burning just below this region is responsible for the C
generation. Meanwhile, the observed enhanced Ne/O and low Fe/O, N/O,
and Mg/O abundance ratios can be explained as a natural consequence of
the s-process within the ``pulse driven convection zone'' (Herwig
2005). The Ne may be predominantly $^{22}$Ne, which can be readily
generated --- at the expense of $^{14}$N --- within the He burning
shell. The $^{22}$Ne can then serve as an iron-depleting neutron
source during the s-process. In such a scenario, one therefore expects
Ne to be enhanced, while Fe, N and Mg are depleted relative to O
--- as observed (Table 3). 

\subsection{The temperature of X-ray-emitting plasma within BD
  +30$^\circ$3639}

 The two plasma temperatures obtained from our best-fit,
  two-component APED model --- $1.7\times10^6$ K and $2.9\times10^6$ K
  --- likely represent the extremes of a continuous range of
  temperature within BD$+30^\circ$3639. Hence the spectral diagnostics
  available in the LETG/ACIS X-ray spectrum of BD$+30^\circ$3639
  demonstrate conclusively, for the first time, the presence of a
  temperature gradient within the X-ray-emitting region of a PN.
  However, even the higher of these two temperatures, obtained from
those line ratios diagnostic of the hottest plasma present in the
LETG/ACIS-S 1st-order spectrum, is far lower than that expected from
simple adiabatic shock models, given the present-day central star fast
wind speed (700 km s$^{-1}$; Leuenhagen et al.\ 1996; Marcolino et
al.\ 2007).  This discrepancy between observed and predicted hot
bubble plasma temperatures has been noted by many investigators over
the past decade (e.g., Arnaud et al. 1996; Chu et al.\ 2001; Soker \&
Kastner 2003; and references therein), and was recently discussed by
Kastner et al.\ (2008) in their analysis of the collective Chandra and
XMM-Newton data compiled to date for PN hot bubbles. The temperature
discrepancy has previously been explained as indicative of heat
conduction from the tenuous hot bubble to the dense, relatively cool
swept-up shell (Stute \& Sahai 2007; Sch\"{o}nberner et al. 2006;
Steffen et al. 2008) or mixing of the two media (Chu et al 1997; Stute
\& Sahai 2006). 

However, such models of mixing and heat conduction predict that the
bulk of the X-ray emission arises in the same gas that is responsible
for the visible-light nebula (Steffen et al. 2008).  Our robust
determination of nonsolar abundances (in particular, greatly enhanced
C and Ne) in the X-ray-emitting plasma within BD$+30^\circ$3639
therefore indicates that the shocked gas is predominantly present-day
stellar wind or jet, as opposed to nebular gas. Hence the heat
conduction and mixing mechanisms, though certainly viable in the
general case, do not play a major role in determining the X-ray
temperature of this particular PN.  We note that this argument likely
holds even at early stages of the heat conduction process. This is
because at early times the conduction goes through the ``evaporation''
stage, where the intermediate-temperature gas comes from the cold
phase (Borkowski et al. 1990). In PNs the cold phase is the visible
nebular gas, and the abundance results obtained here preclude such an
origin for the X-ray-emitting plasma. 

There remains the possibility that the shocked wind presently seen in
X-rays was ejected at an earlier epoch when the fast wind speed was
${\sim}$ 300--400 km s$^{-1}$, a velocity regime more consistent with
the measured range of $T_X$ (Arnaud et al.\ 1996; Akashi et al.\ 2006,
2007). Alternatively, collimated jets, perhaps associated with
molecular ``bullets'' detected in mm-wave interferometric imaging
(Bachiller et al.\ 2000), may be responsible for the X-ray
emission. Such a scenario would be similar to that proposed for NGC
7027 (Kastner et al.\ 2002; Cox et al.\ 2002) and would be
  consistent with the possibility that BD$+30^\circ$3639 is a bipolar
  nebula viewed nearly pole-on (Kastner et al.\ 2002; Lee \& Kwok 2005). 
This possibility will be pursued in our forthcoming
paper concerned with spatial analysis of the LETG/ACIS data (Yu et
al.\ 2008, in preparation).

\section{Conclusions}

Using Chandra's LETG/ACIS-S spectrometer, we have obtained the first
X-ray gratings spectrum of a planetary nebula. The LETG/ACIS-S
spectrum of the young, rapidly evolving BD$+30^\circ$3639 displays
strong emission in the H-like resonance lines of O {\sc viii} and C
{\sc vi} and He-like triplet line complexes of Ne {\sc ix} and O {\sc
  vii}, and appears devoid of lines of highly ionized Fe and N.  Our
spectral modeling, consisting of fits of variable-abundance APED
plasma models with one and two temperature components, demonstrates
that an isothermal plasma is unable to simultaneously
reproduce key spectral features such as the O {\sc vii} to O {\sc
  viii} and Ne {\sc ix} to Ne {\sc x} line ratios and the 6--18 \AA\
continuum. The best-fit two-component plasma model, which is able to
well match these same features, indicates that the X-ray emission line
spectrum is representative of a range of temperatures from $\sim1.7$
MK to $\sim2.9$ MK. These results constitute the first case in
  which a temperature gradient has been inferred within the
  X-ray-emitting region of a PN.

The spectral modeling places tight constraints on the degree
of abundance anomalies present in the X-ray-emitting plasma within
BD$+30^\circ$3639, convincingly demonstrating that Fe is highly deficient
(best-fit Fe/O $\sim0.3$, relative to the solar ratio) and that C and Ne
are highly enhanced in abundance (best-fit ratios C/O $\sim30$ and
Ne/O $\sim4$). This C overabundance, although very large, is not as
pronounced as deduced previously on the basis of X-ray CCD
spectroscopy. In addition, based on the LETG/ACIS-S line spectrum, we
find no evidence for enhanced N and depleted Mg, as inferred
previously from CCD spectra; indeed, if anything, N is somewhat
underabundant in the X-ray-emitting gas. 

The sharply nonsolar composition of the X-ray-emitting plasma is 
similar to that determined for the present-day central star of
BD$+30^\circ$3639 via optical/UV spectroscopy. We conclude that the
plasma consists predominantly of very recently ejected gas originating
from nucleosynthesis processes that occurred deep within the
progenitor AGB star. The ``pristine'' state of this C- and Ne-enriched
(and Fe-depleted) plasma suggests processes such as heat conduction
and/or mixing between the superheated plasma and cooler, denser
nebular gas may not suffice to explain the fact that the inferred
range of X-ray emission temperatures is well below that expected for
shocks generated by the present-day, 700 km s$^{-1}$ central star
wind. Instead, it appears that the shocks detected via X-rays likely
result from lower-speed (300--400 km s$^{-1}$) ejections, perhaps in
the form of collimated jets and/or reflecting the rapid evolution of
the central star wind.  

 The intrinsic X-ray luminosity we deduce from the modeling,
  $\sim10^{33}$ erg s$^{-1}$, is appoximately an order of magnitude
  larger than most previous, CCD-based estimates. This luminosity lies
  at the very highest end of the range of $L_X$ predicted by models
  describing either spherically-symmetric PN hot bubbles or pulsed
  jets in symbiotic stars (e.g., Stute \& Sahai 2006, 2007),
  suggesting that the wind collisions in BD$+30^\circ$3639 are very
  strong indeed. Such strong wind interactions may be commonplace in
  PN with [WC] central stars, however (K08). In these and other
respects, the plasma abundances, temperatures, and luminosity
determined from the dispersed X-ray
spectrum of BD+30$^{\circ}$3639 should serve both to constrain models
of stellar evolution and to guide the development of sophisticated
models of the wind interactions responsible for the superheated gas
within PNs.

\vspace{0.1in} 

\acknowledgments {\it This research was supported by
  NASA through Chandra award GO5--6008X issued to Rochester Institute
  of Technology by the Chandra X-ray Observatory Center, which is
  operated by Smithsonian Astrophysical Observatory for and on behalf
  of NASA under contract NAS8--03060.}




\clearpage




\clearpage

\clearpage

\clearpage

\clearpage

\begin{figure}
\epsscale{.80}
\centering
\includegraphics [scale=0.55, angle=270]{f5a.eps}
\vskip 0.2in
\includegraphics [scale=0.55, angle=270]{f5b.eps}
\caption{As in Fig.~\ref{One_2Vturb_model}, but for the best-fit
  two-component APED model.  \label{Two_comp_model}}
\end{figure}
\clearpage

\clearpage


\clearpage
\input{tab1.tex}

\clearpage
\input{tab2.tex}
\clearpage
\input{tab3.tex}
\clearpage
\input{tab4.tex}

\end{document}

%% file: tab1.tex
\begin{deluxetable}{cccccc}
\tabletypesize{\scriptsize}
\tablewidth{0pt}
\tablecaption{Chandra Observations of BD+30$^{\circ}$3639 \label{tbl-1}}
\tablehead{
\colhead{Obs.ID} & \colhead{Date} & \colhead{Instrument} & \colhead{Exposure (ks)} & \colhead{0$^{th}$ order (counts)} & \colhead{1$^{st}$ order (counts)} \\
\colhead{} & \colhead{} & \colhead{} & \colhead{} & \colhead{source (background)} & \colhead{source (background)} }
\startdata
587 & 2000 Mar 21 & ACIS-S3 & 18.8 & 4762 (238) & N/A\\
\tableline
5409 & 2006 Feb 13 & LETG/ACIS-S & 85.4 & 2039 (964) & 1742 (764)\\
7278 & 2006 Mar 22 & LETG/ACIS-S & 61.8	& 1441 (655) & 1231 (593)\\
\tableline
5410 & 2006 Dec 20 & LETG/ACIS-S & 53.9 & 1187 (572) & 1043 (495)\\
8495 & 2006 Dec 21 & LETG/ACIS-S & 77.1 & 1713 (821) & 1480 (763)\\
8498 & 2006 Dec 24 & LETG/ACIS-S & 19.9 &  511 (214) &  386 (211)\\
\enddata
\tablecomments{Source counts are calculated without subtracting background.}
\end{deluxetable}

%% file: tab2.tex
\begin{deluxetable}{ccccccc}
\tabletypesize{\scriptsize}
\tablewidth{0pt}
\tablecaption{List of line fluxes of BD+30$^{\circ}$3639. \label{tbl-2}}
\tablehead{
\colhead{} & \colhead{${\lambda}_{\it lab}$$^{a}$} & \colhead{${\lambda}_{\it o}$$^{b}$} & \colhead{FWHM$^{c}$}  & \colhead{${\it f}$$_{l}$$^{d}$} \\
\colhead{Line} & \colhead{ (\AA)} & \colhead{ (\AA)} & \colhead{(arcsec)} & \colhead{(10$^{-6}$ photons cm$^{-2}$ s$^{-1}$)}  }
\startdata
Si {\sc xiii} ....................& 6.648      &    6.781  & ......                 &   0.27 [0.06-1.42]       \\
Mg {\sc xii} .................... & 8.419      &    8.401  & ......                 &   0.5  [0.1-1.0]           \\
Mg {\sc xi}   ....................& 9.169      &    9.203  &   4.6 [1.1-20.1]       &   1.1  [...]                \\
                                  & 9.231      &    9.265  &    $''$                &   1.2e-4 [...]             \\
                                  & 9.314      &    9.348  &    $''$                &   0.4  [...]                \\
Ne {\sc ix}   ....................& 11.544     &   11.563  &   3.9 [2.3-5.8]        &   4.8  [2.8-7.4]           \\
Ne {\sc x}    ....................& 12.132     &   12.126  &   2.5 [0.001-3.9]      &   3.8  [2.3-5.7]           \\
Ne {\sc ix}   ....................& 13.447     &   13.457  &   2.9 [2.4-3.4]        &  27.2  [21.6-33.2]         \\
                                  & 13.553     &   13.563  &   $''$                 &   6.2  [0.7-11.2]          \\
                                  & 13.699     &   13.709  &   $''$                 &  16.7  [13.4-20.1]         \\
Fe {\sc xvii} ....................& 15.014     &    .....  & ......                 &    (2.0)$^{e}$       \\
O {\sc viii}  ....................& 15.176     &   15.191  & ......                 &   6.3  [1.8-18.0]          \\
O {\sc viii}  ....................& 16.004     &   15.997  & ......                 &   5.2  [2.7-14.5]          \\
Fe {\sc xvii} ....................& 16.780     &    ...... & ......                 &    (0.5)$^{e}$       \\
Fe {\sc xvii} ....................& 17.051     &    ...... & ......                 &    (0.3)$^{e}$       \\
Fe {\sc xvii} ....................& 17.096     &    ...... & ......                 &    (0.2)$^{e}$       \\
O {\sc vii}   ....................& 18.627     &   18.618  &   3.8 [2.5-5.0]        &  17.9  [13.0-22.1]         \\
O {\sc viii}  ....................& 18.967     &   18.966  &   2.5 [2.1-3.0]        &  40.0  [34.0-46.0]         \\
O {\sc vii}   ....................& 21.602     &   21.607  &   2.8 [2.2-3.3]        &  38.2  [31.1-44.8]         \\
                                  & 21.807     &   21.809  &   $''$                 &  13.1  [7.8-17.8]          \\
                                  & 22.098     &   22.103  &   $''$                 &  27.7  [21.1-34.0]         \\
S {\sc xiv}   ....................& 24.285     &   24.322  & ......                 &   5.1  [0.5-21.8]          \\
N {\sc vii}   ....................& 24.7798    &   ......  & ......                 &   (9.0)$^{e}$              \\
C {\sc vi}    ....................& 26.357     &   26.330  &   2.4 [1.5-3.5]        &  11.2  [4.9-17.3]          \\
                                  & 26.990     &   26.963  &   $''$                 &  22.9  [13.1-33.0]         \\
                                  & 28.465     &   28.438  &   $''$                 &  31.4  [10.0-53.5]         \\
C {\sc vi}    ....................& 33.734     &   33.708  &   2.7 [1.3-4.0]        &  98.0  [79.0-145.0]        \\
C {\sc v}     ....................& 34.973     &   ......  & ......                 &  (55.0)$^{e}$              \\
\enddata
\tablenotetext{a}{Theoretical wavelength of identification, from APED. In 
case of a multiplet, we give the wavelength of the stronger component.}
\tablenotetext{b}{Measured wavelength.}
\tablenotetext{c}{FWHM of emission lines in units of arcsec; see Sec 3.1.}
\tablenotetext{d}{90{\%} confidence intervals are given in brackets.}
\tablenotetext{e}{Flux is a 1 $\sigma$ upper limit. The line 
detected at 15.1911 ${\AA}$ line may be a blend of ${\lambda}$ 15.176 O {\sc viii} and 
${\lambda}$ 15.014 Fe {\sc xvii}, but we estimate that the Fe {\sc xvii}
line contributes no more than $\sim50${\%} of the blended line flux.}
\end{deluxetable}

%% file: tab3.tex
\begin{deluxetable}{ccc}
\tabletypesize{\scriptsize}
\rotate
\tablecaption{BD+30$^\circ3639$: APED model fitting results$^{a}$ \label{tbl-3}}
\tablewidth{0pt}
\tablehead{
\colhead{PARAMETER}            & \colhead{single-component APED model} & \colhead{two-component APED model}  }

\startdata
N$_{H}$ (10$^{22}$ cm$^{-2}$) .& 0.21 [0.19-0.24]       & 0.24    [0.20-0.28]     \\
EM$_{1}$$^{b}$ .................& $2.8\times10^{55}$ [$(2.1-4.0)\times10^{55}$]    & $6.9\times10^{54}$ [$(3.4-13.8)\times10^{54}$] \\
EM$_{2}$$^{b}$ .................&         -               & $1.7\times10^{55}$ [$(1.0-2.2)\times10^{55}$] \\
T$_{1}$ (10$^{6}$ K) .........& 2.3 [2.2-2.4]          & 2.9 [2.6-3.3]           \\
T$_{2}$ (10$^{6}$ K)..........&      -                 & 1.7 [1.3-2.1]           \\
C .........................& 19.50 [12.29-28.82]       & 28.30 [16.05-45.85]       \\
N .........................& 0.17 [0.02-0.39]       & 0.38  [0.08-0.87]        \\
O .........................& 0.44 [0.26-0.67]       & 0.87  [0.57-1.29]        \\
Ne ........................& 2.03 [1.39-2.85]       & 3.30  [2.21-4.16]        \\
Mg .......................& 0.60 [0.23-0.97]       & 0.63  [0.31-1.50]        \\
Fe ........................& 0.13 [0.05-0.23]       & 0.19  [0.10-0.32]        \\
Flux$^{c}$ ....................&   4.3e-12              &   5.0e-12               \\
$L_X$$^{d}$ .....................&   7.4e+32              &   8.6e+32               \\

\enddata
\tablenotetext{a}{Confidence intervals are given in brackets, and all abundances are expressed as ratios to their solar values. See Sec. 3.2.2.}
\tablenotetext{b}{Emission measure in units of cm$^{-3}$.}
\tablenotetext{c}{Intrinsic (unabsorbed, N$_{H}$ = 0) X-ray fluxes in units of ergs cm$^{-2}$ s$^{-1}$.}
\tablenotetext{d}{Intrinsic X-ray luminosity ($D$=1.2 kpc) in units of ergs s$^{-1}$.}
\end{deluxetable}

%% file: tab4.tex
\begin{deluxetable}{lccccccc}
\tabletypesize{\scriptsize}
\rotate
\tablecaption{BD+30$^\circ3639$: comparison of LETG and X-ray CCD
  spectrum plasma model fitting results \label{tbl-3}} 
\tablewidth{0pt}
\tablehead{
\colhead{PARAMETER} &  \colhead{A96} & \colhead{K00} & \colhead{M03} & \colhead{G06} & \colhead{M06} & \colhead{This paper}               \\
\                   &   ASCA/SIS     &        \multicolumn{3}{c}{Chandra/ACIS-S3}    &   Suzaku/XIS  &  Chandra/LETG/ACIS }            
\startdata
N$_{H}$ (10$^{22}$ cm$^{-2}$) .....& 0.12           & 0.1 [0.09-0.11] & 0.24 [0.23-0.25] & 0.2 [1.7-2.3]  & 0.21[1.4-2.5] &    0.24 [0.20-0.28]\\
$L_X$ (10$^{32}$ ergs s$^{-1}$)....& 1.3-1.7        &   2.3          & -             &   -          &     12       &   8.6       \\
T$_{1}$ (10$^{6}$ K) ..............& 3.0  [2.7-3.3] & 2.7 [2.6-2.8]   & 2.1  [2.08-2.12] & 2.4 [2.1-2.6]  & 2.2 [2.1-2.3] &    2.9  [2.6-3.3] \\
T$_{2}$ (10$^{6}$ K) ..............&  -              &  -              & -                &  -             &  -            &    1.7  [1.3-2.1]  \\
C/O ........................&  281            &  -              & 84   [79-90]     & 97  [8.1-813]  & 85  [71-101]  &    32.5 [15-45] \\
N/O ........................&  7.2            &  -              & 2.2  [2.0-2.3]   & 5.8 [1.3-80.1] & 3.2 [0.9-5.5] &    0.4  [0.0-1.0] \\
Ne/O ......................&8.3            &  -              & 4.6  [3.9-5.3]   & 5.4 [0.1-57.7] & 5.8 [4.7-7.5] &    3.8  [3.3-5.0] \\
Mg/O ......................&    -             &  -              & 0.24 [0.22-0.26] & 0.3 [0.1-0.8]  &  -            &    0.7  [0.4-1.5] \\
Fe/O .......................&  0              &  -              & 0                & 0.3 [0.1-0.8]  & ${<}$ 0.1     &    0.2  [0.1-0.4] \\
\enddata
\tablecomments{90{\%} confidence intervals are given in brackets, and
  all abundance ratios are expressed as relative to their solar values. Intrinsic luminosity ($L_X$) is calculated on the basis of $D$=1.2 kpc. \\
References: A96 = Arnaud et al. 1996; K00 = Kastner et al. 2000; M03 =  Maness et al. 2003; G06 = Goergiev et al. 2006; M06 = Murashima et al. 2006}
\end{deluxetable}